\documentclass[aps,prb,showpacs,twocolumn,floats,superscriptaddress,a4paper]{revtex4-1}
\usepackage{amssymb}
\usepackage{amsbsy}
\usepackage{amsmath}
\usepackage{graphicx}
\usepackage{textcomp}
\usepackage[body={18cm,24.3cm},top={2.5cm},left={1.5cm}]{geometry}

\let\la=\label

\def\nn{\nonumber}

\def\bpm{\begin{pmatrix}}
\def\epm{\end{pmatrix}}
\def\be{\begin{equation}}
\def\ee{\end{equation}}
\def\bea{\begin{eqnarray}}
\def\eea{\end{eqnarray}}
\def\ba{\begin{array}}
\def\ea{\end{array}}

\def\etal{\emph{et al. }}
\def\ep{{\epsilon}}

\begin{document}
\title{Localization in coupled heterogeneously disordered transport channels on the Bethe lattice}

\author{Hong-Yi Xie}
\author{M. M\"uller}
\affiliation{The Abdus Salam International Centre for Theoretical Physics, P.O.B. 586, 34100 Trieste, Italy}

\date{\today}

\pacs{72.15.Rn, 72.20.Ee, 72.70.+m, 73.20.Jc}

\begin{abstract}
We study the Anderson localization in systems, in which transport channels with rather different properties are coupled together. 
This problem arises naturally in systems of hybrid particles, such as exciton-polaritons, 
where it is not obvious which transport channel dominates the coupled system. 
Here we address the question of whether the coupling between a strongly and a weakly disordered channel 
will result in localized (insulating) or delocalized (metallic) behavior. Complementing an earlier study in 1D [H.~Y.~Xie, V.~E.~Kravtsov, and M.~M\"uller, Phys. Rev. B \textbf{86} 014205 (2012)], the problem is solved
here on a bilayer Bethe lattice with parametrically different parameters.
The comparison with the analytical solution in 1D shows that dimensionality plays a crucial role. 
In $D=1$ localization is in general dominated by the dirtier channel, which sets the backscattering rate. In contrast, 
on the Bethe lattice a delocalized channel remains almost always delocalized, even when coupled to strongly localized channels.  
We conjecture that this phenomenology holds true for finite dimensions $D>2$ as well. 
Possible implications for interacting many-body systems are discussed.   
\end{abstract}

\maketitle
 
\section{Introduction}  \label{intro}
In a variety of physical contexts, the situation arises that two or more propagating channels with 
different transport properties are coupled together, competing with each other or modifying each other's properties. 
Under these circumstances it is interesting to study the resulting localization properties on the coupled system. 
Such a question arises in particular in the context of exciton-polaritons, which are hybrid particles: half photons, half excitons, the two channels being coupled linearly via dipolar interaction.~\cite{savona07} Another realization of this physical situation can be found in bilayer graphene, with different degrees of disorder affecting the two layers. A recent work proposed such bi- or trilayers as field effect transistors, whereby a gate potential controls the degree of disorder sensed by the electrons in the bilayer.~\cite{Dx} 

Similar questions arise in the problem of energy or matter  
localization in few- or even many-body problems, where a multitude of propagation channels may exist to 
transport particles or energy from one place in the system to another. For example, energy may be transported in small, 
nearly independent units in the form of quasiparticles, or it may have a propagation channel in which a larger amount of 
energy is propagating in the form of blobs of several quasiparticle-like excitations that form 
sorts of bound states. Such ``bound states" were argued to be favorable transport channels 
in the context of few-particle problems. The problem was especially studied in low dimensions,~\cite{Shepelyansky, Imry, Xie2012} 
where under certain circumstances such compounds are found to have an enhanced localization length as compared to single-particle excitations. The question arises, then, as to which channel of propagation is the most favorable in transport problems containing a larger number of particles, or in the situation of particles at finite density.  

In this type of problem, the various propagation channels are not independent, but couple to each other by scattering events. 
Understanding transport in such interacting systems is a challenging and largely unresolved problem. 
Here we do not aim at resolving all aspects of the many-body problem, but address one sub-question which arises in its context.  
Indeed, the interacting systems have a common feature with noninteracting hybrid particles: 
Two or more propagating channels with parametrically different localization properties are coupled together and influence each other's transport characteristics. Under these circumstances it is interesting to study what are the resulting localization properties in the coupled system. In particular in the specific case where a less localized system is coupled to a more localized one, the question arises as to which of the two components eventually dominates the transport: Does one obtain an insulating or conducting system? A central result of our work is to show that the answer to this question depends crucially on the dimensionality of the system. 

In our recent work~\cite{XKM2012} the question of the competition between alternative propagation channels 
was raised in disordered one-dimensional systems. This case can be studied in great detail in the form of a single-particle problem with two parallel, coupled channels. Among others, this naturally describes the Anderson localization of exciton-polaritons in quasi-one-dimensional semiconductor heterostructures. By exactly solving the Anderson model on a two-leg ladder ($D=1$), we found two regimes whose localization properties are qualitatively different: (i) a resonant regime, where the ``slow" chain (the more disordered one) dominates the localization length of the ladder; this can be understood as a manifestation of the fact that in one dimension the backscattering rate determines the localization properties of a coupled system, since in general the localization length is of the order of the mean free path; (ii) an off-resonant regime, where the ``faster" chain helps to delocalize the slow chain, although with low efficiency. 

In that 1D study the disorder was taken to set the smallest energy scale, which allows for a fully analytic solution of the problem.
In higher dimensions ($D > 2$), however, weak disorder has no significant effect on localization. 
Hence, we are restricted to considering intermediate or strong disorder in order to address meaningfully  the question of the 
role of interchannel coupling. Meanwhile, since the disorder is comparable to or stronger than the hopping strengths, 
resonance conditions, as in regime (i) of the weakly disordered 1D chains, are impossible. 
Furthermore, in contrast to the physics in one dimension, proliferation of backscattering plays a subdominant role for the Anderson transition of the coupled system, and therefore, the resulting phenomenology of coupled-channel problems turns out to be rather different.

In this paper we study two coupled Bethe lattices with different transport characteristics. 
This can be viewed as the limit of infinite dimensions ($D \to \infty$) of the problem of coupled channels, which we will contrast with the case of two coupled chains ($D=1$). The behavior on the Bethe lattice is suggestive of the physics 
to be expected in high-dimensional systems. Indeed, we believe that the qualitative behavior of coupled lattices in $D>2$ 
is very similar to the phenomenology found on the Bethe lattice. However, the latter has the significant advantage of being exactly solvable, which we exploit below.

\begin{widetext}
\begin{center}
\begin{figure}
\includegraphics[width=0.85\textwidth]{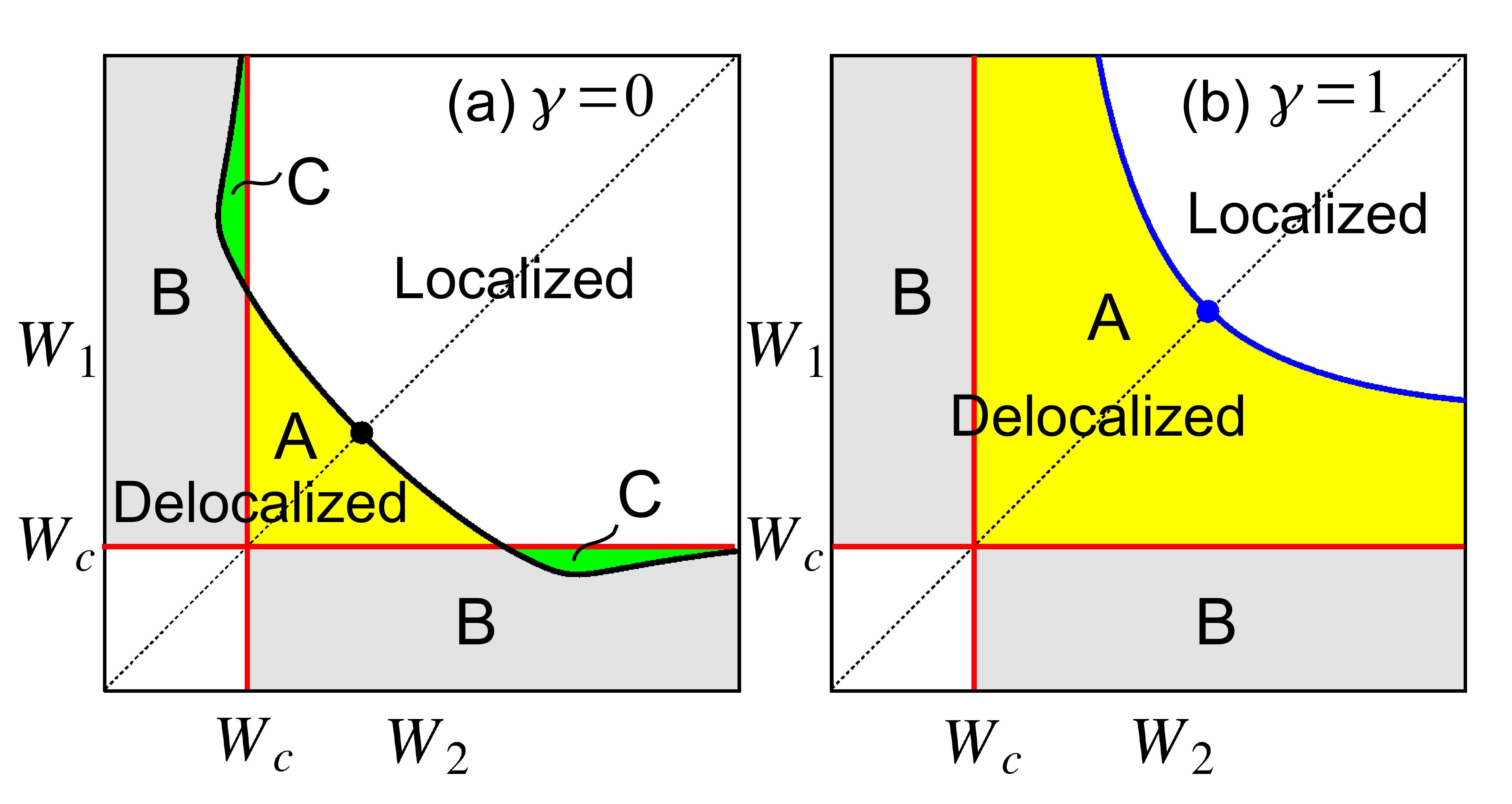}
\caption{Schematic phase diagram for coupled Bethe lattices with identical hopping strength,
but different disorder $W_1\neq W_2$, 
as inferred from the results in Figs.~\ref{equiv-latts} and \ref{gam12-equiv-hop}. The critical disorder for uncoupled lattices, 
$W_{c}\approx 17.3$, is indicated by the red lines. (a) Nearest-neighbor intralattice coupling only, $\gamma=0$. The mobility edge for the middle of the band ($E=0$) of the coupled system is 
indicated by the black curve. In region $A$ (yellow), in the absence of the coupling $t_\perp$ the two lattices would both be localized. However, the finite $t_\perp$ pushes the system into the delocalized phase. 
If in the absence of coupling one lattice is delocalized and the other one localized, there are two possibilities. 
In region $B$ (gray), the coupled system becomes delocalized; that is, the less disordered channel dominates. 
Only when the delocalized lattice is very close to criticality and is coupled to a very strongly disordered lattice [region $C$ (green)] localization prevails. However, this regime occurs in a very narrow window of parameters. 
(b) Next-nearest-neighbor hopping included, $\gamma=1$. The mobility edge is indicated by the blue curve. 
In region $A$ (yellow) the coupling between two localized lattices induces a delocalized phase. 
In contrast to (a), the region $C$ is eliminated by the next-nearest-neighbor hopping: 
The coupled system is \emph{always} more delocalized than either of the uncoupled lattices. 
} \label{bethe-ph-diag}\
\end{figure}
\end{center}
\end{widetext}

Statistical models on the Bethe lattice~\cite{HansBethe, Baxter} have attracted a lot of studies, 
because they admit exact solutions and reflect features of the corresponding systems in sufficiently high 
spatial dimensions. The Anderson model on the Bethe lattice was first introduced and solved 
by Abou-Chacra, Anderson, and Thouless in Refs.~\onlinecite{AAT1} and \onlinecite{AT2}, 
where the existence of the localization transition was proven and the location of the mobility edge was found.  
That work showed in particular that localization is possible in the \emph{absence} of loops in the lattice. 
The model was solved by studying the self-consistency equation for the on-site self-energy, which leads to 
a nonlinear integral equation for the probability distribution function of that quantity. The transition from the localized phase 
to the delocalized phase is characterized by the instability of the fixed point distribution of real self-energies 
with respect to a perturbation with infinitesimal imaginary parts of the self-energies. 
Physically, the latter describes an infinitesimally weak coupling to a dissipative bath which allows for decay processes.  The above instability signals that local excitations start coupling to a bath on sites infinitely far away, which signals their spatial delocalization.
 
The stationary distribution function of the self-energy can be found numerically with the help of a population dynamics, 
or ``pool," method.~\cite{AAT1, MonGarel} The original work by Abou-Chacra \etal~\cite{AAT1, AT2} has inspired a number of 
studies in both the physics~\cite{MirlinFyov9197, MilDer94, MonGarel, BiroliTar, BiroliTar2} and 
the mathematics~\cite{KunzSouillard, AcoKlein, Klein, AizSimWar, AizWar} communities. 
The recent work Ref.~\onlinecite{BiroliTar2} points out that the Anderson model on the Bethe lattice may 
have a further transition within the delocalized phase and corresponds to a transition in the level statistics. 
Here, we focus however on the standard delocalization transition, as discussed by Abou-Chacra \etal

In the present work we generalize the approach by Abou-Chacra \etal to the case of two coupled Bethe lattices.
Following Refs.~\onlinecite{AAT1} and  \onlinecite{AT2} we derive a recursion relation for the local Green's functions 
(encoded in a $2\times2$ matrix in layer space) and study the effect of interlayer coupling on the location of the transition. 
We restrict ourselves to the band center ($E=0$). Furthermore, we focus on the case of lattices with identical hopping, 
but different disorder strengths. This choice is motivated by the one-dimensional case, where equal hoppings lead to resonance effects, which enhance the localization tendency in the coupled system. In contrast, we find that despite the choice of equal 
hoppings such a localizing effect almost never occurs on coupled Bethe lattices. This is illustrated by the schematic phase diagrams of Fig.~\ref{bethe-ph-diag}, which anticipate and summarize the main results of our analysis: 
Under most circumstances the coupling between two layers {\em enhances delocalization}. 
Only when one couples a barely metallic layer to a strongly disordered second layer 
and excludes next-nearest-neighbor interlayer couplings [$\gamma=0$ in the Hamiltonian (\ref{full-ham-bethe}) below], 
the coupling can induce localization. However, in the largest part of the phase diagram the coupling has a delocalizing effect. 
In the case of next-nearest-neighbor interlayer couplings (i.e., nearest-neighbor coupling across layers, $\gamma=1$), 
the coupled layers are always delocalized if one of the uncoupled layers is delocalized. Moreover, in some range of parameters a coupling between two localized lattices can induce delocalization. 

Our central result may be summarized by the statement that on Bethe lattices the delocalized lattice essentially dominates the physics. In other words, if a delocalized channel exists, delocalization, diffusion, and the ability of entropy production will persist even upon coupling to more localized channels. As mentioned before this is quite opposite to the phenomenology in 1D where most often the more disordered chain dominates the localization properties.

The remainder of the paper is organized as follows. 
In Sec.~\ref{bethe-recur} we define the Anderson model 
on two coupled Bethe lattices and derive the recursion relation for the local Green's functions. 
In Sec.~\ref{pop-dym} we present the population dynamics algorithm, 
which is used to study the statistics of the local self-energy. In Sec.~\ref{ph-digrm} 
we obtain the location of mobility edges, which gives rise to the phase diagrams shown in Fig.~\ref{bethe-ph-diag}. 
Their qualitative features will be explained by a perturbative analysis. 
Finally, we discuss the role of dimensionality and the possible implications of our results on the interacting particles in the Conclusion.

\section{Bilayer Anderson model}   \label{bethe-recur}

\subsection{Model}
\begin{center}
\begin{figure}
\includegraphics[width=0.45\textwidth,height=6cm]{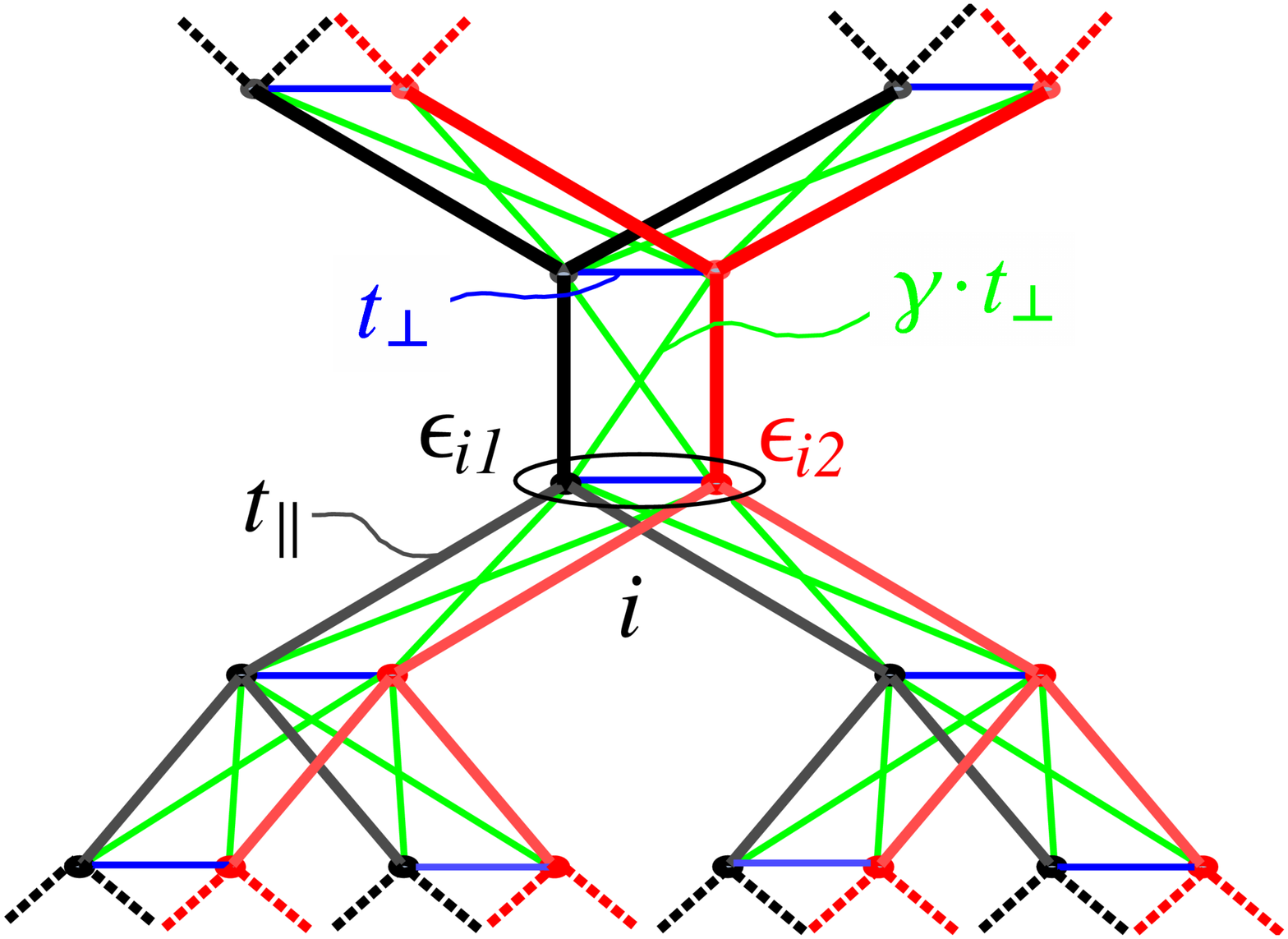}
\caption{Anderson model on a bilayer Bethe lattice, described by the Hamiltonian (\ref{full-ham-bethe}), 
shown for connectivity $K+1=3$. We consider two types of interlayer coupling: 
(i) Only nearest-neighbor coupling (horizontal blue lines), setting $\gamma=0$. 
(ii) Additional next-nearest-neighbor coupling (diagonal green lines), with $\gamma=1$.} \label{bethe-latt}
\end{figure}
\end{center}

We consider two Bethe lattices labeled by $\nu=\{1,2\}$. 
A Bethe lattice is defined as the interior of an infinite regular Cayley tree, each vertex having the same coordination number $K+1$. 
The essential feature of such a lattice is the absence of loops.  
The Bethe lattice can be realized as the thermodynamic limit of a random regular graph of constant connectivity $(K+1)$, that is, a graph where each site connects to other $K+1$ sites, which are randomly and uniformly selected. It is known that any finite portion of such a graph is a tree, with probability tending to one as the size tends to infinity. 
The advantage of the random-graph construction is the explicit absence of  
boundary effects. A random graph can thus be viewed as a regular tree wrapped onto itself.

Analogously to the two-chain model studied in Ref.~\onlinecite{XKM2012},
we define the Anderson model on coupled Bethe lattices as (cf. Fig.~\ref{bethe-latt})
\be     \la{full-ham-bethe}
\begin{split}
& H = \sum_{\nu=1,2}{ \left( {\sum_{i}{\ep_{i \nu} c_{i \nu}^\dagger c_{i \nu}}-
t_{\parallel} \sum_{\langle i,j \rangle}{\left(c_{i \nu}^\dagger c_{j\nu}+h.c.\right)}} \right) } \\
  & -t_{\perp} \left( \sum_{i}{ c_{i 1}^\dagger c_{i 2}} 
     + \gamma \sum_{\langle i,j \rangle}{\left(c_{i 1}^\dagger c_{j 2}+ c_{i 2}^\dagger c_{j 1} \right)} +h.c. \right).
\end{split}
\ee
Here $i$ labels the coordinates of two corresponding sites in the two layers, and $\langle i,j \rangle$ denotes two 
nearest neighbors $i$ and $j$ on the Bethe lattice. We take the onsite energies $\ep_{i\nu}$ to be independently distributed 
random variables with zero mean. $t_\parallel$ is the  nearest-neighbor hopping strength within each layer. 
As motivated above, we take the intralayer hoppings to be equal, so as to come closest to the resonant case in one dimension, 
which shows the strongest localizing effects. $t_\perp$ is the interlayer hopping strength. 
In addition to direct (nearest-neighbor) interlayer coupling, 
we also allow for next-nearest-neighbor hoppings of strength $\gamma t_\perp$. 
We will consider  the two cases $\gamma=0$ and $\gamma=1$. 
For $K=1$, the former reduces to the 1D model studied in Ref.~\onlinecite{XKM2012}. 

The reason to include finite next-nearest-neighbor hoppings is as follows. 
Consider the effect of coupling a first lattice to another one with very strong disorder or vanishing hopping. 
If we exclude next-nearest-neighbor hopping by setting $\gamma=0$, 
the only significant effect of the coupling is to increase the effective disorder on the first lattice, 
while the renormalization of its hopping is strongly subdominant or even absent altogether. However, the latter is not the case 
if we allow for next-nearest-neighbor interlayer hopping. Indeed, this introduces a weak but nonnegligible additional propagation 
channel, which proceeds via the disordered lattice.  As the study below will show (and as anticipated in Fig.~\ref{bethe-ph-diag}), 
in the case $\gamma=1$, the renormalization of the hopping $t_\parallel$ always dominates over the enhancement of effective disorder, and hence, the coupling always has a delocalizing tendency. In the context of the more general problems of coupled parallel propagation channels in many-body systems, the case of $\gamma\neq 0$ appears to be a rather generic and natural choice. Even a rather small $\gamma$ is sufficient to avoid the phenomenology found for $\gamma=0$, which leads to atypical behavior in some small regions of the phase diagram.  

In the Hamiltonian~(\ref{full-ham-bethe}) the two layers are subject to different random potentials, characterized by 
two probability distribution functions $p_\nu(\ep)$. For convenience we assume them to be box distributed: 
\be   \label{rand-pot-dit}
p_\nu(\ep) = \begin{cases} 1/W_\nu, & \ep \in [-W_\nu/2 , W_\nu/2], \\ 
                           0, &  \text{otherwise}. \end{cases}
\ee

Our goal is to study the effect of weak interlayer coupling $t_\perp$ on the Anderson transition of the system. 
As mentioned above, this parallels the case of two resonant chains described in Fig.~7 of Ref.~\onlinecite{XKM2012}. 
However, as we will discuss in detail in Sec.~\ref{ph-digrm}, the parameter range of interest on the Bethe lattice 
is $W_{1,2} \gtrsim t_\perp, t_\parallel$, in contrast to the weak disorder limit considered in Ref.~\onlinecite{XKM2012}. 
However, the notion of resonant interlayer coupling is meaningful only if the disorder is so weak that a well-defined dispersion 
relation exists, which is not the case for the regime of interest on the Bethe lattice. 
Therefore, the equality of the two intralayer hoppings $t_\parallel$ does not have important consequences in the present study.

\subsection{Recursion relation for the local Green's functions}

The retarded Green's function at energy $E$ is defined by
\be
G_{i\mu,j\nu} (E) = \langle i ,\mu | \frac{1}{E + i\eta -\hat{H}} |j,\nu \rangle,
\ee
where $\nu,\mu\in \{1,2\}$ are layer labels, the kets stand for
\be
 |i,\nu \rangle \equiv c_{i \nu}^\dagger |\text{Vacuum}\rangle ,
\ee
and $\eta$ is an infinitesimal positive real number, representing an infinitesimally weak coupling to a dissipative bath into which particles can decay. We introduce $2 \times 2$ matrices in the layer space, $\hat{H}_i$, $\hat{G}_i$, and $\hat{T}$, whose elements are
\begin{subequations}
\be
(\hat{H}_{i})_{\mu\nu}= \langle i ,\mu | \hat{H} |i,\nu \rangle, 
\ee
\be
(\hat{G}_i)_{\mu\nu}= G_{i\mu,i\nu},
\ee
and
\be 
\hat{T}_{\mu\nu}= -\delta_{\mu\nu} t_{\parallel} - (1-\delta_{\mu\nu}) \gamma t_\perp.
\ee
\end{subequations}
$\hat{T}$ describes the hopping from one pair of sites to a neighbor pair of sites.
 
One can easily show that $\hat{G}_i=\hat{G}_i(E)$ satisfies the following equation:
\be  \label{gre-rec-1}
\hat{G}_i = \frac{1}{\displaystyle E+i\eta - \hat{H}_i - \hat{T} \sum_{j \in \partial{i}}{\hat{G}_j^{(i)}} \hat{T}},
\ee 
where $\partial{i}$ denotes the set of neighbors of $i$. The $\hat{G}_j^{(i)}$ are the Green's functions at the coordinate $j$ in the absence of all bonds between the pairs of sites at $i$ and $j$. $\hat{G}_j^{(i)}=\hat{G}_j^{(i)}(E)$ satisfies the recursion relation,
\be  \label{gre-rec-2}
\hat{G}_j^{(i)} = \frac{1}{\displaystyle E+i\eta - \hat{H}_j - \hat{T} \sum_{k \in \partial{j} \setminus i}{\hat{G}_k^{(j)}} \hat{T}},
\ee
where $\partial{j}\setminus i$ denotes the set of neighbors of $j$ excluding $i$. $\hat{G}_i$ and $\hat{G}_j^{(i)}$ are complex symmetric matrices. In order to obtain $\hat{G}_i$, we first solve the recursion relation~(\ref{gre-rec-2}), and then substitute the solution into Eq.~(\ref{gre-rec-1}). 

The self-energies are defined via the layer-diagonal matrix elements $(\hat{G}_{j}^{(i)})_{\nu\nu}$ as
\be   \label{def-self-energy}
S_{j\nu}(E)=E + i\eta - \ep_{j\nu} - 1/(\hat{G}_{j}^{(i)})_{\nu\nu}.
\ee
Their imaginary parts,
\be  \label{im-self-energy}
 \Gamma_{j\nu}(E) \equiv \text{Im}S_{j\nu}(E),
\ee
characterize the decay processes of local excitations overlapping with $|j,\nu \rangle$ and having energy $E$.\cite{Anderson1958}
 
Under the recursion (\ref{gre-rec-2}) the $\Gamma_{j\nu}$ assume a stationary distribution, 
whose characteristics determine whether the system is in the localized phase 
or in the delocalized phase.~\cite{Anderson1958, AAT1, AT2, BiroliTar2} In the thermodynamic limit,
one has 
\be  \label{crt-mit}
\lim_{\eta \to 0}\lim_{\mathcal{N} \to \infty} P(\Gamma_1 > 0\,\,\text{or}\,\, \Gamma_2 > 0) = \begin{cases} 0, \quad \text{localized}, \\ >0, \quad \text{delocalized},  \end{cases}
\ee
$\mathcal{N}$ being the number of lattice sites. The thermodynamic limit, $\mathcal{N} \to \infty$, and the limit of vanishing dissipation, $\eta \to 0$, do not commute, since in a finite system, whose spectrum is discrete, $\eta \to 0$ always leads to vanishing $\Gamma_{j\nu}$'s. Note that the values $\Gamma_{\nu=1,2}$ on the two sublattices are statistically dependent in the presence of coupling; in particular, they are of the same order of magnitude.  

We emphasize that the average value of $\Gamma_{\nu}$, namely $\langle \Gamma_{\nu} \rangle$, \emph{cannot} be used to identify the Anderson transition, because in the localized phase an infinitesimal dissipation $\eta$ leads to long tails in the distribution function of $\Gamma_{\nu}$, which leads to a finite value $\langle \Gamma_{\nu} \rangle$. Instead, one needs to consider the typical value of $\Gamma_{\nu}$, as defined by the geometric average
\be     \label{typ-values}
\Gamma_{\text{typ},\nu} = e^{\langle \ln{\Gamma_\nu} \rangle},
\ee
which depends on the lattice label $\nu$ if the two lattices are statistically not identical. 
However, as they are of the same order of magnitude, the localization transition can be identified by either of the two typical values, by locating the boundary between the two regimes:
\be  \label{crt-mit-2}
\lim_{\eta \to 0}\lim_{\mathcal{N} \to \infty} \Gamma_{\text{typ},\nu} = \begin{cases} 0, \quad \text{localized}, \\ >0, \quad \text{delocalized}.  \end{cases}
\ee
The equivalence of $\Gamma_{\text{typ},\nu=1,2}$ for the purpose of identifying the phase transition will be shown explicitly in Sec.~\ref{pop-dym}, based on the population dynamics.

\section{Anderson transition and population dynamics}  \label{pop-dym}
A convenient way to determine the mobility edge
was proposed in Refs.~\onlinecite{AAT1} and \onlinecite{AT2}. 
It is based on analyzing the stability of the real solution of Eq.~(\ref{gre-rec-2}) at the energy 
$E$ with respect to the insertion of the infinitesimal imaginary energy shift $i\eta$. 
In the localized phase the real solution is stable.
In contrast, in the delocalized phase, the solution develops a finite imaginary part, 
which implies that $\Gamma_{\text{typ},\nu}(E+i\eta) \neq 0$ as $\eta \to 0$. 
The physical interpretation of this criterion is as follows. For a finite but large tree, 
if the boundary sites are coupled to a bath with a dissipation rate $\eta$, we test whether the dissipation 
at the root of the tree, measured by $\Gamma_{\text{typ},\nu}{(E)}$, is vanishing or not as the tree size tends to infinity. 
If $E$ belongs to the localized part of the spectrum (point spectrum), particle transport is absent at large scale and 
there is no dissipation at the root. In contrast, in the delocalized regime, we observe finite dissipation even deep inside the tree. This procedure in fact implements the criterion~(\ref{crt-mit}) for the Anderson transition, as the instability of real self-energies reflects the Anderson
transition as a phenomenon of spontaneous breakdown of unitarity of the scattering matrix associated with the system Hamiltonian.\cite{Fyorov2003}

The stability analysis can be realized by a population dynamics, which is a numerical recipe to solve the stochastic iteration Eq.~(\ref{gre-rec-2}). A detailed description of such an algorithm for the single-lattice case can be found in Refs.~\onlinecite{AAT1} and \onlinecite{MilDer94, MonGarel, BiroliTar, BiroliTar2}. 
The basic idea is to simulate the distribution of a random variable $X$ by the empirical distribution of a large population of representatives $\{X_\alpha\}_{\alpha=1}^{\mathcal{M}}$. Here the random variable $X$ is the symmetric $2\times2$ matrix $\hat{G}_j^{(i)}$. For simplicity, we denote $\hat{G}_j^{(i)}$ by $\hat{G}$, and the population by $\{\hat{G}_\alpha\}_{\alpha=1}^{\mathcal{M}}$. The $\mathcal{M} \gg 1$ representatives can be understood as values of Green's functions on sites at a given distance from the root on a large tree. The population dynamics consists of a number of sweeps of the population, which simulate the propagation of dissipation step by step towards the root of the tree, whereby the number of representatives is kept constant.\cite{MonGarel} At the $n_s^{\rm th}$ stage, we denote the population as $\{\hat{G}_{\alpha,n_s}\}_{\alpha=1}^{\mathcal{M}}$, which are obtained with the following procedure:
 
(i) As an initial condition for the population we chose the Green's functions of $\mathcal{M}$ 
uncoupled sites subject to a random potential and a small dissipation, that is, $\{\hat{G}_{\alpha,0}\}_{\alpha=1}^{\mathcal{M}}$ with matrix elements
\bea
(\hat{G}_{\alpha,0})_{\nu\nu} &=& (E - \ep_{\alpha\nu} +i\eta)^{-1}, \quad \nu \in \{1,2\}, \\
(\hat{G}_{\alpha,0})_{12}&=&(\hat{G}_{\alpha,0})_{21}=0,\nn
\eea
where $\ep_{\alpha\nu}$ are independently drawn from the probability distribution~(\ref{rand-pot-dit}). 
$\eta$ is taken as small positive number, representing the dissipation on the boundary sites of the tree. 

(ii) Generate the $n_s^{\rm th}$ population from the $(n_s-1)^{\rm th}$ population. For each member $\beta = 1,2, \cdots, \mathcal{M}$ of the new population, 
we choose $K$ matrices randomly and uniformly from the population $\{\hat{G}_{\alpha,n_s-1}\}_{\alpha=1}^{\mathcal{M}}$, 
called $\{\hat{G}_{\alpha_1,n_s-1}, \cdots, \hat{G}_{\alpha_K,n_s-1} \}$, and generate $2K$ random numbers according to the probability distribution function in Eq.~(\ref{rand-pot-dit}), 
called $\{ \ep_{1\nu}, \cdots, \ep_{K\nu}\}$ with $\nu =1,2$. Substitute these quantities on the right-hand side of Eq.~(\ref{gre-rec-2}) with $\eta =0$ since the dissipative bath only couples to the boundary sites, 
and obtain $\hat{G}_{\beta,n_s}$ on the left-hand side. 

We calculate the typical value of $\Gamma_\nu$ in the population $\{\hat{G}_{\alpha,n_s}\}_{\alpha=1}^{\mathcal{M}}$,
\be
\ln{\Gamma_{\text{typ},\nu}^{(n_s)}} =\frac{1}{\mathcal{M}} \sum_{\alpha=1}^{\mathcal{M}}{\ln{\Gamma_{\alpha,\nu}^{(n_s)}}},
\ee
using Eqs.~(\ref{def-self-energy}) and (\ref{im-self-energy}). 
The localization transition can be determined by studying the evolution of $\Gamma_{\text{typ},\nu}^{(n_s)}$ under sweeps. 
As shown in Ref.~\onlinecite{AAT1}, if the $\Gamma_{j\nu}$'s are small enough, the recursion relation (\ref{gre-rec-2}) 
leads to a \emph{linear homogeneous} equation for $\Gamma_{j\nu}$, and 
the growth of the typical value of $\Gamma_{j\nu}$ under sweeps is dominated by the largest eigenvalue of the 
linearized recursion relation. Therefore, as long as $\Gamma_{\text{typ},\nu}^{(n_s)}$ is sufficiently small, 
statistically $\Gamma_{\text{typ},\nu}^{(n_s)}$ grows linearly with the growth rate
\be
\lambda_{n_s} = \ln{\Gamma_{\text{typ},\nu}^{(n_s)}} - \ln{\Gamma_{\text{typ},\nu}^{(n_s-1)}}.
\ee
Notice that in this linear regime as long as the two lattices are coupled, the statistics of $\lambda_{n_s}$ is \emph{independent} 
of the lattice index $\nu$. In other words, $\Gamma_{\text{typ},\nu=1,2}^{(n_s)}$ deviate from zero simultaneously as the system crosses into the delocalized phase, and therefore the criterion for delocalization transition (\ref{crt-mit-2}) 
does not depend on $\nu$. The statistical analysis of $\lambda_{n_s}$ below is restricted to the linear regime where the $\Gamma_\nu$ remain small.

The average growth rate of $\Gamma_{\text{typ},\nu}^{(n_s)}$ over $n_s \gg 1$ successive sweeps is given by
\be  \label{grow-r-num}
\overline{\lambda}= \frac{1}{n_s} \sum_{n_s^\prime=1}^{n_s}{\lambda_{n_s^\prime}},
\ee
and the standard deviation is
\be  \label{variance-lam}
\delta\lambda = \sqrt{ \frac{1}{n_s} \sum_{n_s^\prime=1}^{n_s}{(\lambda_{n_s^\prime}-\overline{\lambda}})^2}. 
\ee
Physically, $|\overline{\lambda}|^{-1}$ may be interpreted as a localization length in the insulating phase, or as a correlation length in the delocalized phase. The Anderson transition occurs when\cite{MonGarel} 
\be  \label{crt-mit-3}
\overline{\lambda} = 0.
\ee 

We obtained numerical results using a population size $\mathcal{M}=10^7$, dissipation $\eta =10^{-15}$, and $n_s=200$ sweeps. 
The statistics of $\lambda_{n_s}$ was collected only after about $10$ sweeps to avoid the initial transient. 
We checked that the $\eta$ dependence of $\overline{\lambda}$ and $\delta\lambda$ was very weak, as long as $\eta$ was taken to be small enough.

\section{Phase diagram}  \label{ph-digrm}
Let us now analyze the effect of the interlayer coupling $t_\perp$ on the Anderson transition. 
For convenience we focus on the band center, $E=0$. We concentrate on relatively weak interlayer coupling 
$t_\perp \lesssim t_{\parallel}$, which guarantees that in the absence of disorder the energy bands 
are not substantially changed by the coupling. In this case a mobility edge first appears at the band 
center $E=0$ upon increasing the hopping strength.~\cite{wcoup} Below 
we present the numerical results of the population dynamics, which lead to the phase diagrams shown in Fig.~\ref{bethe-ph-diag}. In Sec.~\ref{anay-two-bethe} a perturbative analysis is given to explain the qualitative features of the phase diagram.

\subsection{Numerical results}  \label{num-two-bethe}

In the numerical calculations we took a connectivity $K+1=3$ and hopping strengths $t_{\perp} =t_\parallel  = 1$. We analyze the two cases in turn.

\subsubsection{Statistically identical lattices ($W_1=W_2$)} \label{stat-id}
\begin{figure}
\centering
\includegraphics[width=0.47\textwidth]{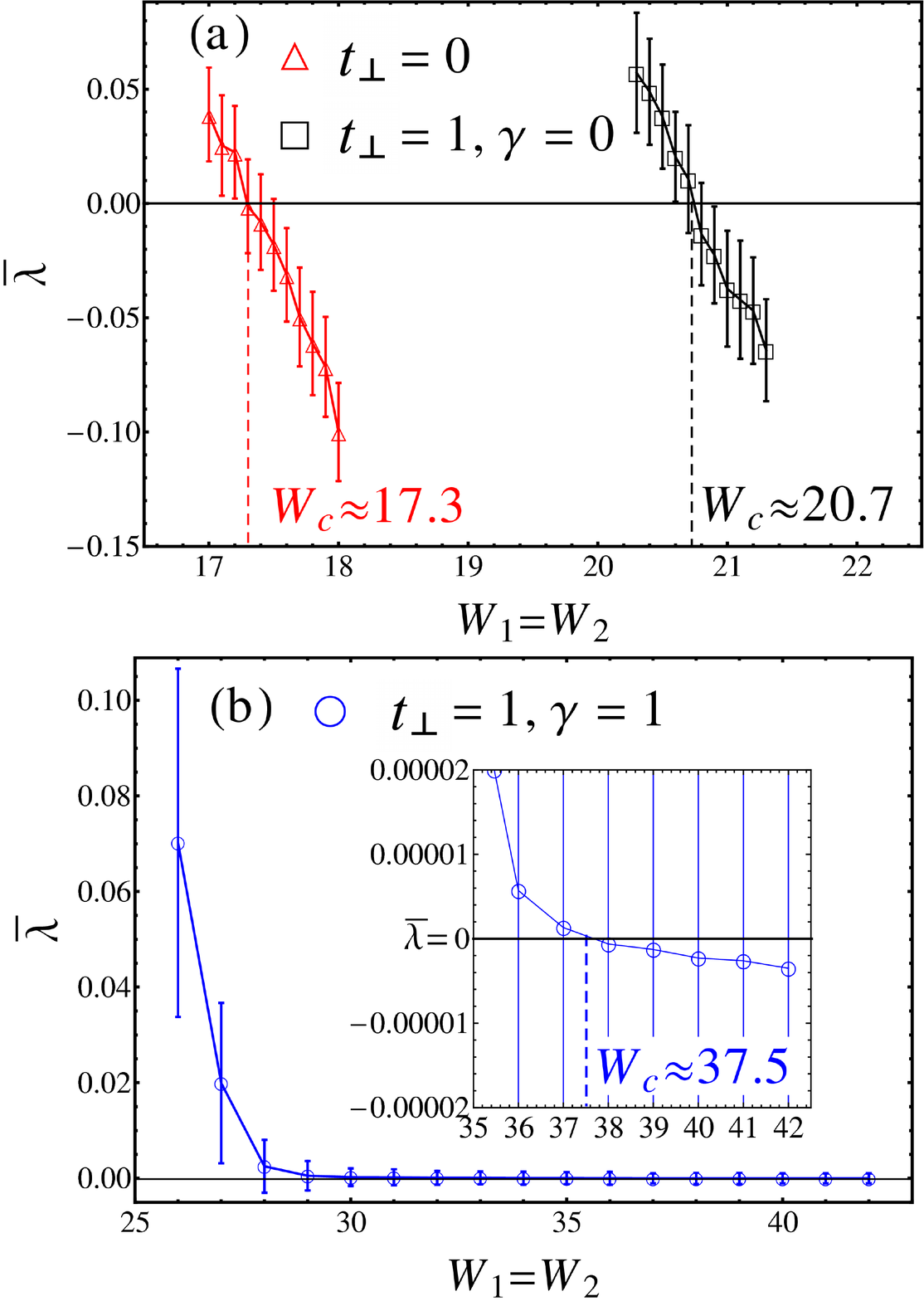}
\caption{Numerical results for the growth rates $\overline{\lambda}$ [Eq.~(\ref{grow-r-num})] 
at the band center for \emph{statistically identical} Bethe lattices as functions of disorder strength $W_1=W_2$. 
Energies are in units of $t_\parallel =1$. The error bars correspond to $\delta{\lambda}$ [Eq.~(\ref{variance-lam})]. 
For the uncoupled lattices (red triangles) the critical disorder strength is $W_c(t_\perp=0) \approx 17.3$. 
Upon coupling the lattices we find the critical disorder strengths: (a) $W_c(t_\perp=1, \gamma=0) \approx 20.7$ (black squares); 
(b) $W_c(t_\perp=1, \gamma=1) \approx 37.5$ (blue circles).} \label{equiv-latts}
\end{figure}

We first analyze two statistically identical lattices with disorder strength 
$W=W_1=W_2$ (following the diagonal line in Fig.~\ref{bethe-ph-diag}). 
In Fig.~\ref{equiv-latts} we show $\overline{\lambda}\pm \delta{\lambda}$ at $E=0$ as functions of the disorder strength 
for uncoupled and coupled lattices. The transition point is determined by Eq.~(\ref{crt-mit-3}). 
For the uncoupled lattices we find the critical disorder $W_c(t_\perp=0) \approx 17.3$, 
which agrees with the results in Refs.~\onlinecite{BiroliTar} and \onlinecite{BiroliTar2}. 
For coupled lattices, the critical disorder strength increases to $W_c(t_\perp=1,\gamma=0) \approx 20.7$ with nearest-neighbor 
coupling only, and to $W_c(t_\perp=1,\gamma=1) \approx 37.5$ when next-nearest-neighbor coupling is included.
Thus the critical disorder is enhanced by the coupling, $W_c(t_\perp \neq 0) > W_c(t_\perp=0)$. 
This implies that if two decoupled lattices are in the localized phase but close enough to criticality, 
the coupling will delocalize the system.

\subsubsection{Parametrically different lattices ($W_1 \neq W_2$)} 

Let us now study two Bethe lattices with identical hopping but different disorder strengths. 
We take $W_1 = W_c(t_\perp=0) \approx 17.3$ to be critical (following the red line $W_1=W_c$ in Fig.~\ref{bethe-ph-diag}) and analyze whether the coupling to a more disordered lattice pushes the system 
to a localized or delocalized phase. If the interlayer coupling is weak $t_\perp \lesssim t_\parallel$, for both $\gamma = 0$ and $\gamma = 1$, we expect that the system is delocalized when $W_2$ is not much larger than $W_c$. This is expected from the results of the preceding section. However, for $W_2 \gg W_c$ the situation may depend on the type of interlayer coupling. 

\begin{figure}
\begin{center}
\includegraphics[width=0.47\textwidth]{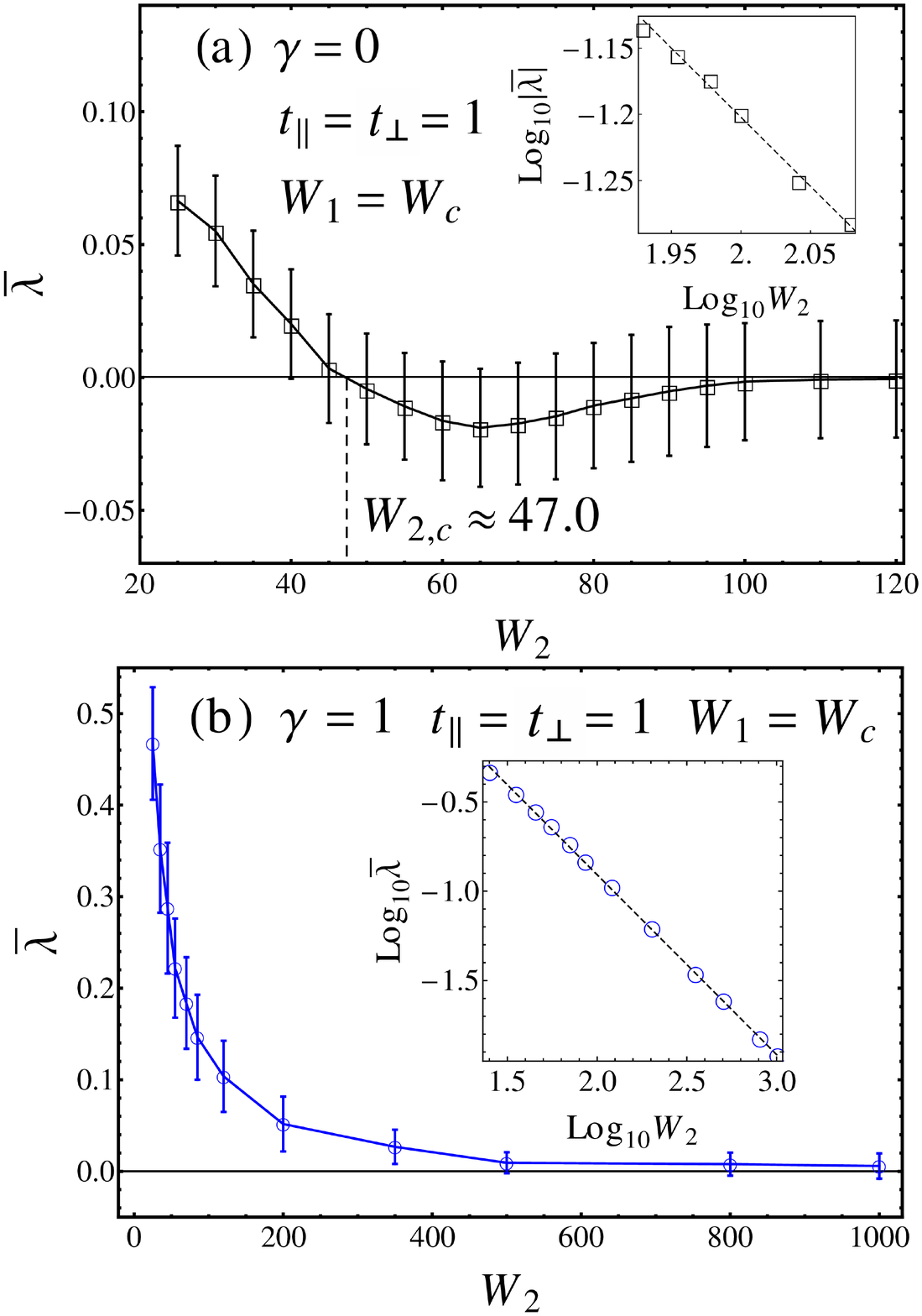}
\caption{
\emph{Statistically nonidentical} Bethe lattices: Numerical results for the growth rate $\overline{\lambda}$ [Eq.~(\ref{grow-r-num})] at the band center as functions of the disorder strength $W_2$. The disorder strength on the $1$ lattice is fixed at $W_1=W_c(t_\perp=0) \approx 17.3$ and the interlayer 
coupling is $t_\perp=1$. The other parameters are the same as in Fig.~\ref{equiv-latts}. (a) For the nearest-neighbor coupling 
there is a mobility edge, $W_{2,c}(t_\perp=1,\gamma=0) \approx 47$. As $W_2 \to \infty$ the system approaches criticality from the
 localized phase, and $\overline{\lambda} \sim -1/W_2$, as expected analytically. 
(b) With next-nearest-neighbor coupling the system is always in the delocalized phase and approaches the transition point 
like $\overline{\lambda} \sim 1/W_2$ as $W_2 \to \infty$ (best fit shown as dashed line in the log-log plot).}       \label{gam12-equiv-hop}\
\end{center}
\end{figure}

In Fig.~\ref{gam12-equiv-hop}, 
we show $\overline{\lambda}\pm \delta{\lambda}$ as functions of $W_2$. We observe the following features:
For $\gamma=0$ a mobility edge occurs at the fairly large disorder $W_2=W_{2,c}(t_\perp=1,\gamma=0) \approx 47$. In other words, as long as $W_2 < W_{2,c}$ the band center becomes delocalized, while it is localized beyond $W_{2,c}$. 
However, when next-nearest-neighbor hopping is included, with relative strength  $\gamma=1$, 
the band center becomes always delocalized upon coupling, for any value of $W_2$.
As $W_2 \to \infty$, the two lattices decouple effectively, and the band center tends back to criticality, 
from the localized and the delocalized side, for $\gamma=0$ and $\gamma=1$, respectively. 
Empirically we find that $\overline{\lambda} \sim  c(\gamma)/W_2$ for large $W_2$, where $c(\gamma =0)<0$ and $c(\gamma =1) >0$. 
As will become clear from the perturbative analysis in Sec.~\ref{anay-two-bethe}, this is due to the suppression or enhancement of the probability of resonances between two neighboring sites on the first lattice. That effect is of 
the order of $1/W_2$.   

The results obtained in Figs.~\ref{equiv-latts} and \ref{gam12-equiv-hop} give rise to the schematic 
phase diagram shown in Fig.~\ref{bethe-ph-diag}. One can distinguish three regions according to the effect of the interlayer coupling: 

(i) Region $A$ (yellow area). In the absence of coupling the two lattices are both localized 
but close enough to criticality. The coupling pushes the two nearly critical lattices into the delocalized phase.    

(ii) Region $B$ (gray area). The better conducting lattice is (sufficiently far) in the delocalized phase, 
while the more disordered lattice is strongly localized. 
The coupled system is nevertheless delocalized due to the dominance of the better channel. 

(iii) Region $C$ (green area). In the absence of coupling the less disordered lattice is delocalized 
but very close to criticality. The more disordered lattice is strongly localized. 
If there is nearest-neighbor coupling only ($\gamma=0$), it pushes the system to the localized phase. 
However, this atypical region is entirely absent if a strong enough next-nearest-neighbor coupling is included ($\gamma=1$).

\subsection{Perturbative analysis}  \label{anay-two-bethe}
The salient features of the phase diagrams shown in Fig.~\ref{bethe-ph-diag} can be understood qualitatively by applying 
a perturbative analysis in the limit $W_2 \gg t_\parallel,t_\perp $. The coupling to the strongly disordered second lattice has two competing effects on the first lattice: 
On the one hand, the hopping strength $t_\parallel$ is effectively enhanced. 
On the other hand, the variance of the on-site energies on the first lattice is effectively enhanced, too. 
If the relative enhancement of the hopping dominates, the coupling tends to delocalize the system. 

To leading order in $1/W_2$, the correction for the hopping strength between nearest-neighbor sites $|i,1\rangle$ and $|j,1\rangle$ is
\be \label{hop-correction}
\delta{t}_{1,ij}(\gamma) \approx  \frac{t_\perp^2}{W_2} \left( \gamma Y_{1,ij} + \frac{t_\parallel}{W_2}Y_{2,ij} \right),
\ee 
where
\be \label{y1y2}
\begin{split}
Y_{1,ij} & = W_2 \left( \frac{1}{E-\ep_{i2}} + \frac{ 1}{ E-\ep_{j2}}\right), \\
Y_{2,ij} & = \displaystyle \frac{W_2^2}{(E-\ep_{i2})(E-\ep_{j2})}.
\end{split}
\ee
Likewise, the correction of the local potential on site $|i,1\rangle$ due to self-energy effects is
\be  \label{dis-correction}
\delta{\ep}_{i1}(\gamma) \approx \frac{t_\perp^2}{W_2}\left[ Y_{3,ij} + \gamma^2 Y_{4,ij} \right],
\ee
where
\be \label{y3y4}
Y_{3,ij}  = \frac{W_2}{E-\ep_{i2}}, \quad Y_{4,ij}  =  \sum_{j \in \partial{i}}{\frac{W_2}{E-\ep_{j2}}}.
\ee
$Y_{1,2,3,4}$ are dimensionless random variables whose probability distributions have long tails.\cite{Anderson1958, AltGef} 
Hence both $\delta{t}_{1,ij}$ and $\delta{\ep}_{i1}$ are dominated by rare, large values. 
This implies that the dominant events are those where either the hopping strength or the disorder strength is strongly enhanced, that is, a link is either strongly favored or blocked. The ratio of the probabilities of such enhancements determines which effect is dominant. Notice that a typical value of $\delta{t}_{1,ij}$ is of order $O(1/W_2^2)$ for $\gamma=0$, but $O(1/W_2)$ for $\gamma>0$, 
while $\delta{\ep}_{i1}$ scales as $O(1/W_2)$ regardless of the value of $\gamma$. Therefore, the enhancement of disorder is dominant when $\gamma=0$, that is, when the nearest-neighbor interchain hopping is suppressed.  

Let us now discuss the Anderson transition at $E=0$. We base this discussion on two observations: 
First, the delocalization of wavefunctions on the Bethe lattice has recently been shown to occur along single paths.~\cite{BiroliTar2} We should therefore study the decay rate of excitations along the best possible path for propagation. 
To obtain a qualitative understanding of the effects of coupling, we approximate the propagation amplitude between two remote sites of the first lattice as the product
\be  \label{prod}
A_L = \prod_{i=1}^{L}{\frac{t_{\parallel}+\delta{t}_{1,ij}(\gamma)}{\ep_{i1}+\delta{\ep}_{i1}(\gamma)}}
   = R_L \prod_{i=1}^{L}{\frac{t_{\parallel}}{\ep_{i1}}}, \quad L \gg 1,
\ee
where $L$ is the distance between the two sites. $\prod_{i=1}^{L}{t_{\parallel}/\ep_{i1}}$ is the amplitude in the absence of coupling, and
\be  \label{eff-ratio}
R_L = \prod_{i=1}^{L}{\frac{1+\delta{t}_{1,ij}(\gamma)/t_\parallel}{1+\delta{\ep}_{i1}(\gamma)/\ep_{i1}}},
\ee
represents the enhancement due to the coupling to the second lattice. The amplitude $A_L$ is the lowest order term in an expansion in the hopping. It corresponds to Anderson's ``upper limit'' approximation, which neglects self-energy effects from sites lateral to the considered paths, as well as the regularization of resonances due to higher order corrections from hoppings along the path. 
Based on this approximation one obtains a simple approximate criterion for localization: 
Consider the probability $P_{|A_L| \gtrsim 1}$ that the propagation amplitude $|A_L|$ along the most favorable path exceeds some fixed finite value $O(1)$. 
Localization obtains so long as this probability vanishes in the thermodynamic limit, $P_{|A_L| \gtrsim 1}\to 0$ as $L\to \infty$.~\cite{Anderson1958,AltGef} 

In order to understand the phase diagram in Fig.~\ref{bethe-ph-diag}, let us consider critical disorder on the first lattice, $W_1=W_c$, and study how the probability $P_{|A_L| \gtrsim 1}(W_2)$ depends on $W_2$ via the correction factor $R_L$.
For $W_2\to \infty$, $R_L\to 1$ (in probability), and $P_{|A_L| \gtrsim 1}(W_2 \to \infty)$ behaves critically; that is, it does not decay exponentially with $L$. For finite but large $W_2$, we need to estimate the correction factor $R_L$.
As mentioned above, it is dominated by the rare events in which either the hopping strength or the local disorder are strongly enhanced, such that one of the factors in Eq.~(\ref{eff-ratio}) is significantly different from unity. 

The probability of a strong enhancement of hopping,  $P_{|\delta{t}_{1,ij}| \gtrsim t_\parallel}$, scales with $W_2$ like~\cite{AltGef}
\be  \label{prob-hop}
P_{|\delta{t}_{1,ij}| \gtrsim t_\parallel} \sim \begin{cases} \displaystyle c_0 t_\perp^2 \frac{\ln{W_2}}{W_2^2}, \quad \gamma=0 \\ 
\displaystyle c_1 \frac{\gamma t_\perp^2}{t_\parallel}\frac{1}{W_2}, \quad \gamma >0,
\end{cases}   
\ee
where $c_{0,1} = O(1)$ do not depend crucially on the parameters of the system. The probability of the strong enhancement of local disorder behaves like
\be \label{prob-dis}
P_{|\delta{\ep}_{i1}|\gtrsim |\ep_{i1}|} \sim c_2(\gamma) \frac{t_\perp^2}{W_1} \frac{1}{W_2},
\ee
where $c_2(\gamma) = O(1)$ depends on $\gamma$, and has a finite limit $c_2(\gamma\to 0) > 0$. 

Among the $L$ factors of $R_L$ a fraction of order $P_{|\delta{t}_{1,ij}| \gtrsim t_\parallel}$ is significantly larger than unity, 
and a fraction of order $P_{|\delta{\ep}_{i1}|\gtrsim |\ep_{i1}|}$ terms significantly smaller than unity. Therefore, it is reasonable to assume that a typical value of $R_L$ 
takes the form 
\be \label{type-enhance}
R_{L,\text{typ}} \sim e^{s(\gamma,W_2) L},
\ee
where the Lyapunov exponent $s(\gamma,W_2)$ is
\be  \label{lyapu}
s(\gamma,W_2) =   \alpha P_{|\delta{t}_{1,ij}| \gtrsim t_\parallel} - \beta P_{|\delta{\ep}_{i1}|\gtrsim |\ep_{i1}|},
\ee
with $\alpha, \beta$ of order $O(1)$. Substituting Eqs.~(\ref{prob-hop}) and (\ref{prob-dis}) in Eq.~(\ref{lyapu}), we predict the scaling
\be
s(\gamma,W_2) \sim \frac{c(\gamma)}{W_2}, \quad W_2 \gg W_1,t_\parallel,
\ee
with a coefficient 
\be  \label{c-gamma}
c(\gamma) \sim \frac{t_\perp^2}{t_\parallel} \left( f(\gamma) - \frac{t_\parallel}{W_1} \right),
\ee
where $f(\gamma) \propto \gamma$ for $\gamma \ll 1$. 
Obviously, the condition $c(\gamma) =0$ marks the transition between enhanced and suppressed propagation.
Close to that criticality, the inverse localization or correlation length follows from the growth rate $|\overline{\lambda}|$ [cf. Eq.~(\ref{grow-r-num})] 
which is proportional to $s(\gamma,W_2)$,
\be \label{loc-length}
\overline{\lambda} \propto s(\gamma,W_2)\sim \frac{1}{W_2}.
\ee
The scaling with $1/W_2$ is clearly observed in the numerical data of Fig.~\ref{gam12-equiv-hop}, confirming the dominance of rare events.
  
Let us now discuss the $\gamma$ dependence of $c(\gamma)$. 
Without next-nearest-neighbor interlayer hopping, we have $c(\gamma=0) \approx -7.5< 0$, as we numerically obtain in Fig.~\ref{gam12-equiv-hop}(a). This is due to the fact that 
$P_{|\delta{t}_{1,ij}| \gtrsim t_\parallel}$ is parametrically smaller than 
$P_{|\delta{\ep}_{i1}|\gtrsim |\ep_{i1}|}$ for $W_2 \to \infty$ [cf. Eqs.~(\ref{prob-hop}) and (\ref{prob-dis})]. 
Therefore, for large enough $W_2 (> W_{2,c})$ the more disordered lattice drives a less disordered, critical lattice to the localized phase, as seen in regime $C$ of Fig.~\ref{bethe-ph-diag} (a).
  
However, when $\gamma > 0$, the probabilities for significant corrections $\delta{t}_{1,ij}$ and $\delta{\ep}_{i1}$ 
both scale as $1/W_2$. For large enough $\gamma$, $c(\gamma)$ becomes positive, as one may anticipate from 
Eq.~(\ref{c-gamma}), considering that ${t_\parallel}/{W_1}$ is numerically small at criticality.
Indeed, the case $\gamma=1$ shown in Fig.~\ref{bethe-ph-diag}(b) is already deep in this regime, with $c(\gamma =1)\approx 12.8>0$ [cf. Fig.~\ref{gam12-equiv-hop}(b)].

The equation $c(\gamma=\gamma_c)=0$ has a solution for some $0< \gamma_c<1$. $\gamma_c$ determines the minimal next-nearest-neighbor interlayer hopping which assures delocalization upon coupling to a disordered lattice even in the limit $W_2\gg W_1$.
A naive linear interpolation between  $c(\gamma=0)$ and $c(\gamma=1)$ allows us to obtain a rough estimate
\be
\gamma_c\approx 0.37 
\ee
for the Bethe lattices of connectivity $K+1=3$ considered here.

\section{Discussion and Conclusion}
We have studied the Anderson localization problem on two coupled Bethe lattices, 
which represents a two-channel problem in the limit of infinite dimensions.
Our main result is the finding that a conducting transport channel is hardly ever localized by the coupling to 
more disordered channels. Rather, transport is usually enhanced by such a coupling. 
This holds true except in the case where three conditions are met simultaneously: 
(i) the conducting channel is very close to criticality;
(ii) it is coupled to a strongly localized channel; (iii) next-nearest-neighbor interlayer couplings are strongly suppressed or absent.  Only in these exceptional cases the coupling to localized channels may induce a localized phase in an otherwise conducting channel. The coupling between moderately localized channels may instead induce delocalization.
We believe that these trends persist also in high but finite dimensions ($D>2$) where the metal-insulator transition takes places at strong disorder. This conjecture is based on the observation that in higher dimensions, as well as on the Bethe lattice, delocalization is mostly driven by a sufficiently strong forward scattering, whereas weak localization effects and enhanced backscattering play a much less important role than in $D\leq 2$. We believe that this difference is at the root of the very different phenomenology between coupled Bethe lattices and 1D chains.

In two dimensions, the localization length becomes parametrically larger than the mean-free path at weak disorder. 
However, since the proliferation of weak localization and backscattering leads to complete localization (in the absence of special symmetries), we expect that a well propagating channel becomes more strongly localized upon resonant coupling to a more disordered channel, similarly as in one dimension. It might be interesting to investigate this numerically. Apart from its theoretical interest, the physics of coupled, unequally disordered 2D lattices might also have practical applications. For example, it was recently proposed~\cite{Dx} that a sheet of bilayer graphene with different disorder strength on the two layers could be operated as a 
field effect transistor, whereby  a perpendicular gate bias tunes the effective disorder of carriers.

The study of localization properties of few- or many-particle systems is more subtle than the toy problem which it motivated here in part. The reason is that multiple (much more than two) coupled channels with complicated substructures 
may exist to transport particles or energy. In particular, for few-particle problems, it has been shown that some of the channels, in which a large number of quasiparticle-like excitations are propagating in the form of ``bound states,'' can be more efficient for 
transport than others, in which the excitations propagate essentially as independent units.~\cite{Shepelyansky, Imry, Xie2012} 
This suggests that one might have to think of the many-body problem as having a hierarchy of channels with parametrically different transport properties. It is reasonable to assume that the effective dimensionality of such channels should be the same as that of the system. Our analysis of a two-channel model in 1D has demonstrated that a ``bad'' channel dominates only when
it is resonantly coupled to a better (``faster'') channel. If the two channels are far from resonance, or if they live in higher effective dimensionality, the fast channel almost always dominates the localization properties, as the present study suggests. Moreover, even in 1D resonance conditions are not met very easily. It either requires two channels with equal hopping strength, or an energy close to the band center or the band edges. Summarizing these considerations, we come to the qualitative conclusion that, apart from some exceptional cases, better conducting channels generically dominate the delocalization: A diffusing channel is difficult to shut down by coupling to dirtier channels.
  
In the context of interacting many-particle systems the above leads to the following conjecture: 
In order to establish that a many-particle system conducts and is not fully localized, a 
sufficient condition will be found by identifying the best transport channel and showing that it is delocalized. 
Indeed, our study suggests that the inclusion of coupling to other channels usually only enhances transport. 
This observation should be a central ingredient when generalizing the ideas of Refs.~\onlinecite{Shepelyansky} and \onlinecite{Imry}
to the analysis of quantum dynamics and transport of systems with several particles.    
However, at this stage, the application to many-particle systems remains a conjecture which needs to be tested further. 
For example, one should establish whether coupling to a much larger number of slow channels does not alter our qualitative findings of ``the survival of the fastest".

\begin{acknowledgments}
We are grateful to D. Basko, P. Brouwer, and V. E. Kravtsov for stimulating discussions.
\end{acknowledgments}


\begin{thebibliography}{99}
\bibitem{savona07} V. Savona, J. Phys.: Condens. Matter \textbf{19}, 295208 (2007).
\bibitem{Dx} D. Xue, H. Liu, V. Sacksteder IV, J. Song, H. Jiang, Q. F. Sun and X. C. Xie, J. Phys.: Condens. Matter \textbf{25}, 105303 (2013).
\bibitem{XKM2012} H.~Y.~Xie, V.~E.~Kravtsov, and M.~M\"uller, Phys. Rev. B \textbf{86}, 014205 (2012).
\bibitem{Shepelyansky} D.~L.~Shepelyansky, Phys. Rev. Lett. \textbf{73}, 2607 (1994).
\bibitem{Imry} Y.~Imry, Europhys. Lett. \textbf{30} (7), 405 (1995).
\bibitem{Xie2012} H.~Y.~Xie, \emph{Anderson localization in disordered systems with competing channels} 
(LAP Lambert Academic Publishing, Saarbr\"ucken, 2012), Chap.~2, where aspects of the few-particle problem are discussed. 
It is suggested there that the most efficient transport channel follows a hierarchical structure in the spatial arrangement of the particles.
\bibitem{HansBethe} H.~A.~Bethe, Proc. R. Soc. A \textbf{150}, 552 (1935).
\bibitem{Baxter} R.~J.~Baxter, \emph{Exactly solved models in statistical mechanics} (Academic Press, London, 1982).
\bibitem{Anderson1958} P.~W.~Anderson, Phys.~Rev. \textbf{109}, 1492 (1958). 
\bibitem{AAT1} R.~Abou-Chacra, P.~W.~Anderson, and D.~J.~Thouless, J.~Phys. C \textbf{6}, 1734 (1973).
\bibitem{AT2} R.~Abou-Chacra and D.~J.~Thouless, J.~Phys. C \textbf{7}, 65 (1973).
\bibitem{MirlinFyov9197} A.~D.~Mirlin and Y.~V.~Fyodorov, Nucl. Phys. B \textbf{366}, 507 (1991); Phys. Rev. B \textbf{56}, 13393 (1997).
\bibitem{MilDer94} J.~D.~Miller and B.~Derrida, J.~Stat.~Phys.~\textbf{75}, 357 (1994).
\bibitem{MonGarel} C.~Monthus and T.~Garel, J.~Phys.~A \textbf{42}, 075002 (2009).
\bibitem{BiroliTar}G.~Biroli, G.~Semerjian, and M.~Tarzia, Prog. Theor. Phys. Suppl. \textbf{184}, 187 (2010).
\bibitem{BiroliTar2}G.~Biroli, A.~C.~Ribeiro-Teixeira, and M.~Tarzia, arXiv: 1211.7334 [cond-mat.dis-nn].
\bibitem{KunzSouillard} H.~Kunz and B.~Souillard, J. Physique Lett. \textbf{44}, L411 (1983).
\bibitem{AcoKlein} V.~Acosta and A.~Klein, J. Stat. Phys. \textbf{69}, 277 (1992).
\bibitem{Klein} A.~Klein, Comm. Math. Phys. \textbf{177}, 755 (1996); Adv. Math. \textbf{133}, 163 (1998).
\bibitem{AizSimWar} M.~Aizenman, R.~Sims, and S.~Warzel, Prob. Theor. Rel. Fields \textbf{136}, 363 (2006); 
Comm. Math. Phys. \textbf{264}, 371 (2006).
\bibitem{AizWar} M.~Aizenman and S.~Warzel, Phys.~Rev.~Lett. \textbf{106}, 136804 (2011).
\bibitem{Fyorov2003} Y.~V.~Fyodorov, Pis’ma Zh. Eksp. Teor. Fiz. \textbf{78}, 286 (2003) [JETP Lett. \text{78}, 250 (2003)]. 
\bibitem{wcoup} As in the two-chain problem in Ref.~\onlinecite{XKM2012}, if $t_{\perp}$ becomes very strong as compared to $t_\parallel$ we reach the one-channel regime, where there is a gap between the two clean subbands, and thus only one propagating channel at a given energy. In this case the energy with the largest localization length is no longer at the band center 
(cf. Figs. 6 and 11 of Ref.~~\onlinecite{XKM2012}). A similar situation is expected for coupled Bethe lattices: 
If the coupling is too strong, the mobility edge first appears at some energy $E\neq 0$.
\bibitem{AltGef} D. J. Thouless, J. Phys. C: Solid St. Phys. \textbf{3}, 1559 (1970); B. L. Altshuler, Y. Gefen, A. Kamenev, and L. S. Levitov, Phys. Rev. Lett. \textbf{78}, 2803 (1997). 
\end{thebibliography}
\end{document}